\documentclass{article}
\usepackage{graphicx} 
\usepackage[a4paper,left=3cm,right=3cm,top=3.5cm,bottom=3.5cm]{geometry}
\usepackage{url} 
\usepackage[inline]{trackchanges} 
\usepackage{soul}
\usepackage{apacite}
\usepackage{makecell} 

\usepackage{amsmath}
\usepackage{amssymb}

\title{Improve the estimate of the b-value in regional catalogs by means of the the b-more positive method}

\author{
    E. Lippiello\textsuperscript{1},
    C. Godano\textsuperscript{1}, 
    G. Petrillo\textsuperscript{2}\\[1ex]
    \textsuperscript{1}\small Department of Mathematics and Physics, Università della Campania ``L. Vanvitelli'', Italy\\
    \textsuperscript{2}\small Earth Observatory of Singapore, Nanyang Technological University, Singapore\\[1ex]    
}
\date{}

\begin{document}
\maketitle

\begin{abstract}
The b-value, which controls the slope of the frequency–magnitude distribution of earthquakes, is a critical parameter in seismic forecasting. However, accurately measuring the true b-value is challenging due to the temporal and spatial variations in the completeness of instrumental seismic catalogs. In this study, we systematically compare traditional methods for estimating the b-value with newer approaches, specifically focusing on the b-more-positive estimator based on positive magnitude difference statistics. We conduct this comparison using both synthetic ETAS catalogs, with artificially introduced incompleteness, and instrumental catalogs from five regions: Japan, Italy, Southern California, Northern California, and New Zealand. Our results from synthetic ETAS catalogs reveal that traditional estimators tend to underestimate the b-value, while the b-more-positive estimator provides a more accurate measurement. Similar patterns are observed in instrumental catalogs, suggesting that traditional methods may also underestimate the true b-value in real datasets.
\end{abstract}

\section{Introduction}

The Gutenberg-Richter (GR) law is a fundamental assumption in most current methods for analyzing earthquake data. It provides essential information, such as the seismicity rate across all event magnitudes in a region, quantified by the exponent of the frequency-magnitude distribution, commonly referred to as the $b$-value, which typically takes value around $1$ for shallow tectonic events (\cite{FD93}).

More specifically, the GR law states (\cite{GR44}) that the probability density $p(m)$ of observing an earthquake with magnitude $m$ is given by
\begin{equation}
p(m) = \beta e^{-\beta (m - m_L)},
\label{GR}
\end{equation}
where $\beta$ is the scaling parameter, $m_L$ is a minimum reference magnitude, and the $b$-value is calculated as $b = \beta / \ln(10)$.

Variations in the $b$-value, both spatial and temporal, have been proposed as indicators of several factors influencing seismic regions, including the stress regime, tectonic characteristics, material heterogeneities, and temperature (\cite{TEWW15,WW02,SWW04}). Laboratory experiments, in particular, have demonstrated an inverse relation between the $b$-value and differential stress (\cite{Sch15}). These findings suggest that the $b$-value may serve as a stress proxy, potentially aiding in identifying high-stress zones where significant future earthquakes are more likely to occur (\cite{GW19,GWV20}).

Estimating the $b$-value presents challenges due to the breakdown of exponential decay at lower magnitudes. This issue arises when cataloging instruments cannot fully capture all earthquakes above background noise, establishing a detection threshold magnitude $M_c$—where only earthquakes above $M_c$ are detected with $100\%$ probability. However, $M_c$ can fluctuate due to network limitations, such as uneven spatial coverage or sensor variability (\cite{SW08, MWWCW11, MW12}). Additionally, coda waves from previous, larger events can obscure smaller earthquakes, further influencing $M_c$ and complicating detection (\cite{Kag04, HKJ06, PVIH07, LCGPK16, Hai16, Hai16a, dAGL18, PLLR20, Hai21}). Failure to properly address these factors may lead to a substantial underestimation of the $b$-value.

To address the issue of incomplete data reporting, a common approach is to restrict the calculation of the $b$-value to magnitudes above a chosen threshold $M_{th}$, typically set higher than the completeness magnitude $M_c$. The $b$-value can then be estimated from Eq.(\ref{GR}), yielding (\cite{Aki65})
\begin{equation}
\beta_0 = \frac{1}{\langle m \rangle - M_{th}},
\label{beta0}
\end{equation}
where $\langle m \rangle$ represents the average magnitude above $M_{th}$. Consequently, the $b$-value is given by $b= \beta_0 / \ln(10)$.

An innovative approach to this issue was recently introduced by \cite{VdE21} with the development of the 'b-positive' estimator. This method estimates the $b$-value by analyzing the distribution of magnitude differences, \(\delta m = m_{i+1} - m_i\), between consecutive earthquakes in the catalog. Specifically, for a complete dataset that obeys the GR law (Eq.(\ref{GR})), it can be demonstrated that the distribution of \(\delta m\), denoted as \(p(\delta m)\), follows an exponential form with a coefficient \(\beta_{+} = \beta\). The key insight from \cite{VdE21}, validated through extensive numerical simulations, is that when restricting the analysis to positive values of \(\delta m\), the distribution \(p(\delta m)\) is significantly less sensitive to detection issues compared to \(p(m)\). \cite{LP24} has subsequently demonstrated that accurately identifying the correct \(b\)-value from positive magnitude difference statistics necessitates the consideration of additional conditions. Specifically, one can examine the magnitude differences between pairs of earthquakes that are not necessarily consecutive in the catalog. However, it is essential that their epicentral distances are sufficiently small to ensure similar behavior of \(M_c\) at their respective epicenters. Under these conditions, which form the basis of the 'b-more-positive' estimator, \(p(\delta m)\) is expected to exhibit an exponential decay with a coefficient \(\beta_{++} \simeq \beta\), even in the presence of substantial detection issues.

In this study, we will present a detailed comparison of the $b$-value estimates derived from traditional estimators (Eq.(\ref{beta0})) with those obtained using the $b$-positive estimators, \(\beta_+\) and \(\beta_{++}\). We will first conduct this comparison using synthetic ETAS catalogs, where incompleteness is artificially introduced, as outlined in \cite{PL20,PL23}. The analysis of these synthetic catalogs will provide insights into the behavior of \(\beta_0\), \(\beta_+\), and \(\beta_{++}\) when applied to instrumental catalogs from various geographical regions.

\section{Magnitude incompleteness}\label{Sec_Inc}

Incomplete earthquake catalogs result primarily from two key factors: seismic network density incompleteness (SNDI) and short-term aftershock incompleteness (STAI). SNDI  occurs when it is difficult to detect earthquakes due to a low signal-to-noise ratio. Several factors contribute to this issue, including noise filtering capabilities and, more importantly, the spatial distribution of seismic stations (\cite{MW12}). STAI  arises from detection limitations in the aftermath of large earthquakes. Smaller aftershocks are often obscured by coda waves generated from prior, larger earthquakes (\cite{LPGTPK19}). Empirical observations suggest that STAI can be characterized by a completeness magnitude that decreases logarithmically as a function of the time elapsed since the mainshock (\cite{Kag04, HKJ06}).

Together, SNDI and STAI result in a completeness magnitude $M_c(t_i, \vec{x_i}, {\cal H}_i)$, which depends on the occurrence time $t_i$, the epicenter coordinates $\vec{x_i}$, and the seismic history ${\cal H}_i$ encompassing all previous earthquakes up to time $t_i$. All earthquakes with magnitudes above $M_c$ are reliably recorded in the catalog but
even at the same location $\vec{x_i}$ and for a given seismic history ${\cal H}_i$, $M_c(t_i, \vec{x_i}, {\cal H}_i)$ is not uniquely determined. Factors such as diurnal and seasonal variations, staffing changes, and other external influences introduce fluctuations of the order $\sigma$ around $M_c$. These fluctuations are typically smaller than the overall spatial and temporal variability of $M_c(t_i, \vec{x_i}, {\cal H}_i)$. The presence of a finite $\sigma$ affects the detection function $\Phi\left(m -  M_c(t_i, \vec{x_i}, {\cal H}_i)\right)$, which denotes the probability that an earthquake with magnitude $m$, occurring at time $t_i$ and location $\vec{x_i}$, is recorded in the catalog. This detection function can be expressed as:

\begin{equation}
  \Phi\left(m - M_c\left(t_i, \vec{x_i}, {\cal H}_i\right)\right)
  = \frac{1}{2} + \frac{1}{2} \text{Erf} \left( \frac{m - M_c}{\sigma} \right),
  \label{Phi}
\end{equation}

where $\text{Erf}(y)$ represents the error function. According to this definition on average, approximately $98\%$ of earthquakes with $m \ge M_c + 2\sigma$ are reported.

A number of technical papers provide tools to identify the optimal value of $M_{th}$ such as $M_{th}>M_c$ (\cite{MW12}). In this study, we focus on three methods: the Maximum Curvature (MAXC) technique (\cite{WW00}), the $b$-value stability (MBS) method (\cite{CG02,WW05}), and the CV method (\cite{GPL24,GTPC24}). In the MAXC method, $M_{th}$ is defined as the magnitude bin with the highest frequency of events in the non-cumulative frequency-magnitude distribution.
The MBS method operates on the assumption that the $b$-value approaches its true value and remains constant for  $M_{th} > M_c$, forming a plateau.
In the CV method, $M_{th}$ is identified as the smallest threshold for which the coefficient of variation (the ratio of the standard deviation to the mean value of magnitudes $m$) exceeds 0.93. These three methods can yield different estimates for $M_{th}$, and thus for $\beta$ derived from Eq.(\ref{beta0}). The resulting values of $\beta$ based on these thresholds will be denoted as $\beta_{MAXC}$, $\beta_{MBS}$, and $\beta_{CV}$, respectively.

\section{Data and Methods}

\subsection{Data}
We consider intrumental catalogs from $5$ different seismic regions: Japan, Italy, Southern California, Norther California and New Zeland. For all the catalogs we restrict the analysis to shallow earthquakes (depth smaller than $50$ kms) and magnitudes $m>0$. For the Japan seismicity we consider the JMA catalog after 2000 restricting to the mainland. For the Italian seismicity we consider the Horus catalog (\cite{LRVG20}) after 2005, still restricting to the mainland. For Southern California we consider the relocated catalog (\cite{HYS12}) from January 1981 to March 2022, for Northern  California we consider the relocated catalog (\cite{WS08}) from January 1984 to December 2021. Finally for New Zealand we take the catalog after 2000. The number of $m>0$ earthquakes in the five catalogs is reported in Table \ref{tab2}.

\subsection{The b-more-positive estimator}\label{sec2.1}

It is straightforward to show that in the case of two independent random variables, $m_i$ and $m_j$, distributed according to an exponential law, as in the GR law Eq.(\ref{GR}), the difference $\delta m = |m_i - m_j|$ between any pair of magnitudes $m_i$ and $m_j$ also follows an exponential distribution:

\begin{equation}
p(\delta m) = \frac{1}{2} \beta e^{-\beta \delta m},
\label{pdm1}
\end{equation}

where the factor $\frac{1}{2}$ serves as a normalization term, accounting for both possible cases, $m_i > m_j$ and $m_j > m_i$. In the following, we define the quantity

\begin{equation}
\beta_+ = \left( \frac{1}{N_+} \sum_{\substack{i=1 \\ m_{j} > m_i + \delta M_{\text{th}}}}^{N_+} \left(m_{j} - m_i - \delta M_{\text{th}}\right) \right)^{-1},
\label{beta+}
\end{equation}
where $j=i+1$ and the sum is restricted only to the $N_+$ earthquakes $m_i$ that are followed by an earthquake with $m_{i+1} > m_i + \delta M_{\text{th}}$. 
Maximizing the likelihood, under the assumption that Eq.(\ref{pdm1}) holds (\cite{VdE21,TiGa24}), shows that $\beta_+ = \beta$ up to statistical fluctuations $\delta b$ of the order of $1/\sqrt{N_+}$.

In the presence of incompleteness, magnitudes no longer obey the GR law, and Eq.(\ref{pdm1}) must be replaced by the following expression (\cite{LP24}):

\begin{equation}
  p(\delta m) \propto \beta e^{-\beta \delta m} \Phi\left(m_i + \delta m - M_c(t_j, \vec{x}_j, {\cal H}_j \,|\, m_i)\right) \Phi\left(m_i - M_c(t_i, \vec{x}_i, {\cal H}_i)\right),
\label{pfac3}
\end{equation}

where $\Phi\left(m_i - M_c(t_i, \vec{x}_i, {\cal H}_i)\right)$ is the detection probability as defined in Eq.(\ref{Phi}), and \\ $\Phi\left(m_i + \delta m - M_c(t_j, \vec{x}_j, {\cal H}_j \,|\, m_i)\right)$ represents the detection probability at time $t_j$ and location $\vec{x}_j$, conditional on the previous earthquake $m_i$ being identified and recorded in the catalog. The key insight is that, when the following condition holds:
\begin{equation}
\Phi\left(m_j - M_c(t_j, \vec{x}_j, {\cal H}_j \,|\, m_i)\right) = 1,
\label{cond2}
\end{equation}
then $p(\delta m)$ obeys a pure exponential law with decay controlled by the coefficient $\beta$, even if $\Phi\left(m_i - M_c(t_i, \vec{x}_i, {\cal H}_i)\right) \ll 1$. Two hypotheses are sufficient to ensure the validity of condition (\ref{cond2}):

\begin{itemize}
    \item \textbf{Hypothesis i):}
    \begin{equation}
    M_c(t_j, \vec{x}_j, {\cal H}_j) \le \min \{ m_i, M_c(t_i, \vec{x}_i, {\cal H}_i) \},
    \end{equation}
    for $t_j > t_i$.
    
    \item \textbf{Hypothesis ii):}
    \begin{equation}
    M_c(t_j, \vec{x}_i, {\cal H}_j) \le M_c(t_j, \vec{x}_j, {\cal H}_j).
    \end{equation}
\end{itemize}

The central observation is that, since $m_i$ has been recorded at time $t_i$, if both hypotheses hold,
after imposing the constraint $m_j > m_i +\delta M_{th}$, when $\delta M_{th} \gtrsim 2\sigma$, we can be reasonably confident that $m_j \gtrsim M_c(t_j, \vec{x}_j, {\cal H}_j) + 2\sigma$, and Eq.(\ref{cond2}) is satisfied with high probability.

Hypothesis (i) is easily satisfied if no earthquake with magnitude larger than $m_i$ occurs within the temporal window $(t_i, t_j)$ between the two earthquakes. Specifically, according to STAI, the obscuration effect caused by earthquakes occurring at times $t < t_i$ diminishes in relevance by the time $t_j > t_i$. Additionally, if all earthquakes in the interval $(t_i, t_j)$ have magnitudes smaller than $m_i$, none of these events can cause a completeness magnitude larger than $m_i$. More precisely, hypothesis (i) is also met if any $m > m_i$ earthquakes occur within $(t_i, t_j)$ but are sufficiently distant in space so as not to produce obscuration effects at $\vec{x}_j$.

For hypothesis (ii) to hold, it is necessary that the epicentral distance $d_{ij}$ between $\vec{x}_i$ and $\vec{x}_j$ is small enough that the two earthquakes occur in a region with very similar network coverage. Additionally, $d_{ij}$ must be sufficiently small for previous large earthquakes to have comparable obscuration effects at both epicentral positions $\vec{x}_i$ and $\vec{x}_j$. Consequently, for hypothesis (ii) to be valid, an upper bound $d_{R}$ on $d_{ij}$ must be imposed. The optimal value of $d_R$ depends on the specific spatial configuration of seismic stations and can be associated with the typical distance over which the completeness magnitude $M_c$ can reasonably be considered constant.

Summarizing, in the b-more-positive estimator, we can define a quantity similarly to Eq.(\ref{beta+}):
\begin{equation}
\beta_{++} = \left( \frac{1}{N_{++}} \sum_{\substack{i=1 \\ m_{j} > m_i + \delta M_{\text{th}} \, \& \, d_{ij} < d_R}}^{N_{++}} \left(m_{j} - m_i - \delta M_{\text{th}}\right) \right)^{-1}
\label{beta++}
\end{equation}
where $j$ is the index of the closest subsequent earthquake in time that has magnitude $m_j > m_i + \delta M_{\text{th}}$ and an epicentral distance $d_{ij} < d_R$. The sum extends over all $N_{++}$ earthquakes $i$ for which all events recorded in the interval $(t_i, t_j)$ and within an epicentral distance smaller than $d_R$ from $\vec{x}_i$ have magnitudes $m < m_i$. According to the previous arguments, for $\delta M_{\text{th}} > 2\sigma$, we have that $\beta_{++} \simeq \beta$ even in the presence of incomplete datasets.

The algorithm implementing the b-more-positive estimator $\beta_{++}$ presents the following scheme:
\begin{enumerate}
    \item For all earthquakes $i$ in the catalog:
    \begin{enumerate}
        \item Consider all subsequent earthquakes $j$ until you find an earthquake such that:
        $m_j > m_i$  \quad \text{and} \quad $d_{ij}<d_R$; 
        \item If $m_j \geq m_i + \delta M_{\text{th}}$, evaluate:
        $\sum_i (m_j - m_i)$
        and move to the subsequent $i$;
        \item If $m_i < m_j < m_i + \delta M_{\text{th}}$, move to the subsequent $i$.
    \end{enumerate}
\end{enumerate}
The b-positive estimator represents a specific case of the b-more-positive estimator in which the index $j$ considers only the subsequent earthquake, specifically $j = i + 1$, with the condition $d_R = \infty$.

\section{Results}
\subsection{The evaluation of the b-value in synthetic ETAS catalogs}

\begin{figure*}
\vskip+0.5cm
\noindent\includegraphics[width=15cm]{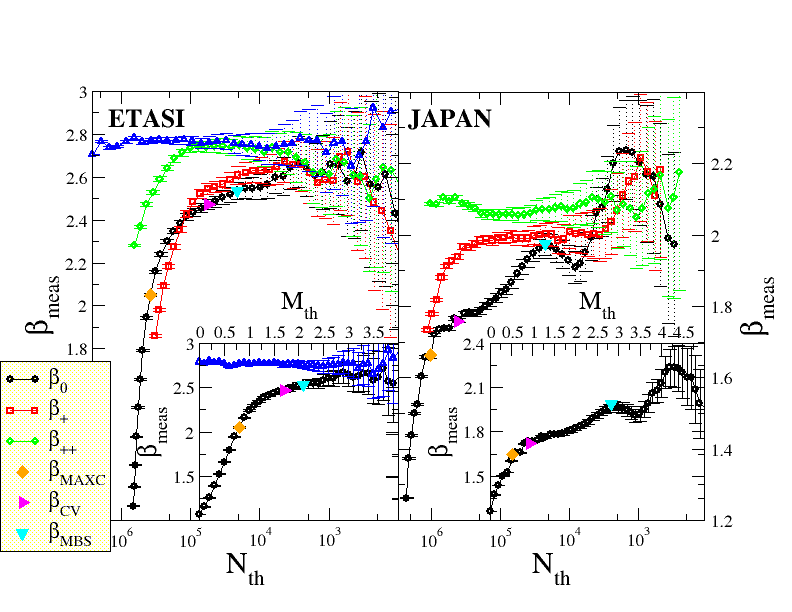}
\caption{(Main Panels) The value of $\beta$ measured using the three estimators, $\beta_0$, $\beta_+$, and $\beta_{++}$, is shown as a function of $N_{th}$ in a synthetic ETASI catalog (main left panel) and in the JMA catalog for Japan (main right panel). 
  Black, red, and green symbols are used for $\beta_0$, $\beta_+$, and $\beta_{++}$, respectively. The values for $\beta_{MAXC}$, $\beta_{CV}$, and $\beta_{MBS}$ are marked by a filled orange diamond, a magenta right triangle, and a cyan downward triangle, respectively.
  Error bars represent twice the standard deviation. For clarity, only in left panels  blue upward triangles denote results from the $\beta_0$ estimator applied to the complete ETAS catalog. 
  (Insets) $\beta_0$ for the ETASI catalog (left panel) and for the JMA catalog (right panel), along with the three estimates $\beta_{MAXC}$, $\beta_{CV}$, and $\beta_{MBS}$, is plotted as a function of $M_{th}$. In the inset of the left panel  we also plot results for $\beta_0$ for the complete ETAS catalog.}
\label{fig1}
\end{figure*}

The Epidemic-Type Aftershock Sequence (ETAS) model effectively captures the main characteristics of the spatio-temporal evolution of seismicity (\cite{Oga88, Oga98}) and is widely regarded as the benchmark for earthquake forecasting (\cite{LoMa10, NOSW19}). 

The key assumption of the ETAS model is that magnitudes are independent variables drawn from the GR law  with a $\beta$ value, which we denote as $\beta_{\text{true}}$. Deviations from the GR law, caused by catalog incompleteness, can be incorporated into the model following the scheme outlined in \cite{dAGL18, PL20, PL23}. More precisely, STAI can be implemented by considering a constant blind time $\tau$ (\cite{Hai16, Hai16a, Hai21}). For any earthquake $j$ in the original complete catalog, we consider all events with magnitude $m_i$ that occurred in the previous time interval $t_i \in (t_j - \tau, t_j)$ and within an epicentral distance $d_{ij} < 50$ km. Each event $j$ is then removed from the catalog with a probability $\Phi(m_i - m_j)$, where $\Phi(x)$ is defined in Eq.(\ref{Phi}) with $\sigma = 0.4$. It is possible to show that the presence of a constant blind time leads to a completeness magnitude decreasing logarithmically in time consistently with instrumental observations (\cite{Hai16,Hai16a}).

To incorporate SNDI, we use the spatial dependence of the completeness magnitude estimated in \cite{SW08} based on the structure of seismic stations in Southern California. We indicate with $m_r(x)$ this local detection threshold, and then further remove events from the original catalog with a probability $\Phi(m_i - m_r(x) + 1) $, where $ \Phi(x) $ is still defined in Eq.(\ref{Phi}) with $ \sigma = 0.4 $.

We refer to the ETAS incomplete model as the ETASI model and generate synthetic ETASI catalogs containing earthquakes with $m_i > 0$, over a period of 25 years. The key quantity in our analysis is $\beta_{\text{true}}$, and we consider different values of $\beta_{\text{true}} \in [2,3]$. For each value of $ \beta_{\text{true}} $, we generate up to ten different synthetic catalogs, each obtained by implementing a different random seed. The values of the other parameters in the ETASI model are not relevant to the present study, and we adopt those parameters optimized for Southern California, as detailed in \cite{PL20}.

We now focus on results for a $b$-value of $b = 1.2$, corresponding to $\beta = 2.763$, though similar observations hold for other choices of $\beta$. We always start from an original dataset containing $N_T \in [5 \times 10^6, 1 \times 10^7]$ earthquakes with $m_i > 0$. After applying STAI, about $70\%-80\%$ of earthquakes are removed. Another significant removal is obtained to implement SNDI, obtaining finals ETASI catalogs with $N_T \in [3 \times 10^5, 9 \times 10^5]$ $m>0$ events, which is comparable with the number observed in instrumental catalogs.

In Fig.\ref{fig1}, we plot results for a specific synthetic ETAS catalog, which initially containes $9961297$ $m>0$ earthquakes, reduced to $2142912$ after STAI and to $668284$ after SNDI; similar patterns are observed for other synthetic catalogs. We show results for the quantity $\beta_0$ evaluated in the original complete catalog, which provides the reference curve for the true $\beta$ value, $\beta_{\text{true}}$. This value is compared with the values of $\beta$ ($\beta_{\text{meas}}$) measured in the final incomplete catalog,
obtained using the three estimators $\beta_0$ (Eq.\ref{beta0}) for different values of $M_{\text{th}}$, $\beta_+$ (Eq.\ref{beta+}), and $\beta_{++}$ (Eq.\ref{beta++}), both evaluated at different values of $\delta M_{\text{th}}$. By increasing $M_{\text{th}}$ and $\delta M_{\text{th}}$, we reduce the number of earthquakes $N_{th}$ used for the three estimators. More precisely, $N_{th}$ represents the number of earthquakes with $m > M_{\text{th}}$ in the evaluation of $\beta_0$ (Eq.\ref{beta0}), $N_{th}=N_+$ in the evaluation of $\beta_+$ (Eq.\ref{beta+}), and $N_{th}=N_{++}$ in the evaluation of $\beta_{++}$ (Eq.\ref{beta++}).
We plot the data as a function of $N_{th}$ to allow a direct comparison among the different estimators. Furthermore, since the standard deviation $\delta \beta$ in all three estimators is approximately $\beta_{\text{meas}} / \sqrt{N_{th}}$, this plot allows us to identify the best estimator as the one that, for fixed $N_{th}$, better approximates $\beta_{\text{true}}$. More specifically, we use $\delta \beta$ for the standard deviation in each estimator, obtained by \cite{TiMu86} for the b-value and by \cite{TiGa24} for the b-positive estimator. In the figure, we display $2\delta \beta$ as error bars to represent this uncertainty.

In the inset of Fig.\ref{fig1} (left panel), we plot $\beta_0$ for both the complete and incomplete catalogs as a function of $M_{\text{th}}$, finding that $\beta_0$ significantly underestimates $\beta_{\text{true}}$ in the incomplete catalog. In fact, in the incomplete catalog, $\beta_0$ grows as a function of $M_{\text{th}}$ but remains consistently below $\beta_{\text{true}}$. A reasonable estimate of $\beta_{\text{true}}$ can only be achieved at very large values of $M_{\text{th}}$, where, however, the uncertainty becomes substantial ($\delta \beta \approx 0.3$) due to the reduced number $N_{th}$. We emphasize that all three estimates of $\beta$ — $\beta_{\text{MAXC}}$, $\beta_{\text{MBS}}$, and $\beta_{\text{CV}}$ — yield values significantly smaller than $\beta_{\text{true}}$. This observation holds not only for the specific synthetic ETAS catalog shown in Fig.\ref{fig1} but also for the other nine synthetic ETAS catalogs generated with the same $\beta_{\text{true}}$ value as well as for other synthetic catalogs with different $\beta_{\text{true}}$ values. 

Fig.\ref{fig1} also shows that $\beta_+$ underestimates $\beta_{\text{true}}$ across all values of $N_+$, except at very large $\delta M_{\text{th}} > 3.8$, where $N_+ \approx 100$ and fluctuations become large enough to encompass $\beta_{\text{true}}$. Interestingly, Fig.\ref{fig1} reveals that $\beta_{++}$ provides a sufficently accurate estimate of $\beta_{\text{true}}$ already for $N_{++} > 10^5$ (when $\delta M_{\text{th}} = 0.8$). This confirms that the $b$-more-positive estimator is highly effective at capturing the true $b$-value, even in the presence of incompleteness. More specifically, we observe that $\beta_{++}$ initially underestimates the true $\beta$ value when $\delta M_{\text{th}} = 0$, then increases monotonically up to $\delta M_{\text{th}} = 0.8$, and subsequently remains approximately constant for larger values of $\delta M_{\text{th}}$.

Following the rationale behind the MBS methodology, we can define the best estimate $\beta^*_{++}$ from $\beta_{++}$ as the value obtained from Eq.(\ref{beta++}) for the first value of $\delta M_{\text{th}}$ such that $\Delta \beta = | \beta_{\text{ave}} - \beta_{++} | \leq \delta \beta_{++}$. Here, $\delta \beta_{++}$ represents the standard deviation of $\beta_{++}$, and $\beta_{\text{ave}}$ is the average value of $\beta_{++}$ over five successive cutoff magnitudes $\delta M_{\text{th}}$, with $\delta M_{\text{th}}$ incremented by $0.1$ in each step. Results for $\beta^*_{++}$ are provided in Tab.\ref{tab2} for the $10$ synthetic ETASI catalogs, showing substantial agreement with $\beta_{\text{true}}$ despite a slight underestimation.

\subsection{The evaluation of the b-value regional catalogs}

\subsubsection{The b-value in the JMA catalog}

Results for the JMA catalog are plotted in the right panel of Fig.\ref{fig1}. The behavior of $\beta_0$ is similar to that observed in the ETASI catalog (left panel). Specifically, we see that $\beta_0$ increases with $M_{th}$, reaching an initial plateau at $\beta_{CV} = 1.74 \pm 0.01$. With further increases in $M_{th}$, $\beta_0$ resumes its growth, eventually stabilizing at a value close to $\beta_{MBS} = 1.96 \pm 0.05$. For even larger $M_{th}$ values, $\beta_0$ shows additional growth toward $\beta \gtrsim 2.1$, albeit with an uncertainty of approximately $\delta \beta \simeq 0.15$.

In contrast, the quantity $\beta_+$ quickly stabilizes at a plateau value around $\beta_{MBS}$ and remains nearly constant across all values of $\delta M_{th}$. Meanwhile, $\beta_{++}$ transitions from an initial value of $\beta_{++} = 2.06 \pm 0.01$ at $\delta M_{th} = 0$ to $\beta_{++} = 2.11 \pm 0.01$ at $\delta M_{th} = 0.4$, after which it decreases slightly to approximately $\beta_{++} = 2.07 \pm 0.01$.

The comparison with Fig.\ref{fig1} highlights that the value of $\beta$ estimated from $\beta_0$ tends to be underestimated. Specifically, we find $\beta_{MBS} = 1.96 \pm 0.05$, which should be compared to $\beta^*_{++} = 2.11 \pm 0.01$. This latter value, $\beta^*_{++}$, is expected to provide the most accurate estimate of $\beta_{true}$.

\subsubsection{The b-value in the Italian, Southern, Northern and New Zekand catalogs}

\begin{figure}
\vskip+0.5cm
\noindent\includegraphics[width=15cm]{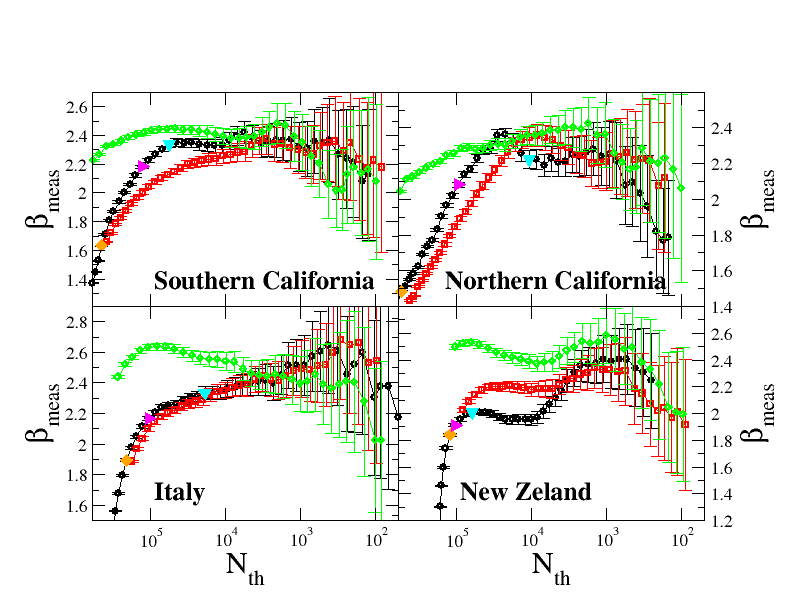}
\caption{The values of $\beta$, measured using the three estimators $\beta_0$, $\beta_+$, and $\beta_{++}$, are plotted as functions of $N_{th}$ for the four geographic regions:  Southern California (upper left panel), Northern California (upper right panel), Italy (lower left panel) and New Zealand (lower right panel). Error bars represent twice the standard deviation, and the same colors and symbols as in Fig.\ref{fig1}, also for the quantities  $\beta_{MAXC}$, $\beta_{CV}$, and $\beta_{MBS}$, are used for consistency.}
\label{fig3}
\end{figure}

The values of $\beta_0$, $\beta_+$, and $\beta_{++}$ as functions of $N_{th}$ for the remaining four geographic regions are shown in Fig. \ref{fig3}. In all cases, $\beta_0$ starts from very low values at large $N_{th}$ and eventually stabilizes in an intermediate regime, roughly constant, meeting the criteria necessary for identifying $M_c$ in both the CV and MBS methods.

As $N_{th}$ decreases further, distinct behaviors emerge across the regions: in Italy, $\beta_0$ continues to increase gradually; in Southern California, it remains approximately stable; while in Northern California and New Zealand, it appears to decline. It is worth noting that in this low-$N_{th}$ regime, statistical fluctuations become significant, and the observed trends may simply reflect random variability rather than a meaningful pattern.

In all four regions, we consistently find that $\beta_0$ is significantly lower than $\beta_{++}$ at large $N_{th}$, aligning with findings from the synthetic ETASI catalog and the results previously observed for Japan (Fig. \ref{fig1}). Fig. \ref{fig3} thus indicates that, also across these additional regions, the traditional estimator $\beta_0$ underestimates the true $\beta$ value. The values for $\beta_{MAXC}$, $\beta_{MBS}$, $\beta_{CV}$, and $\beta^*_{++}$, along with their associated uncertainties, are presented in Table \ref{tab2} for all five geographic regions.

\begin{table}[h!]
\centering
\begin{tabular}{|l|c|c|c|c|c|}
\hline
\textbf{Catalog} & \makecell{\textbf{Num of} \\ \textbf{$m>0$ EQs}} & $\boldsymbol{\beta_{MAXC}}$ & $\boldsymbol{\beta_{CV}}$ & $\boldsymbol{\beta_{MBS}}$ & $\boldsymbol{\beta^*_{++}}$ \\
\hline
\multicolumn{6}{|c|}{\textbf{Instrumental Catalogs}} \\
\hline
Japan & 2313836 & $1.658 \pm 0.006$ & $1.74 \pm 0.01$ & $1.96 \pm 0.05$ & $2.11 \pm 0.01$ \\
Southern California & 794838 & $1.63 \pm 0.01$ & $2.18 \pm 0.02$ & $2.34 \pm 0.04$ & $2.44 \pm 0.04$ \\
Northern California & 856142 & $1.48 \pm 0.01$ & $2.08 \pm 0.03$ & $2.22 \pm 0.09$ & $2.36 \pm 0.09$ \\
Italy & 363936 & $1.89 \pm 0.02$ & $2.16 \pm 0.03$ & $2.33 \pm 0.08$ & $2.61 \pm 0.03$ \\
New Zealand & 165714 & $1.84 \pm 0.02$ & $1.91 \pm 0.02$ & $2.01 \pm 0.03$ & $2.47 \pm 0.05$ \\
\hline
\multicolumn{6}{|c|}{\textbf{Synthetic ETASI catalogs}}
\\
\hline
ETASI 1  & 668284 & $1.986 \pm 0.006$  & $2.078 \pm 0.006$  & $2.53 \pm 0.04$  & $2.71 \pm 0.02$  \\
ETASI 2  & 641320 & $1.942 \pm 0.006$  & $2.036 \pm 0.006$  & $2.57 \pm 0.05$  & $2.73 \pm 0.02$   \\
ETASI 3  & 722492 & $2.103 \pm 0.006$  & $2.178 \pm 0.008$   & $2.55 \pm 0.03$  & $2.74 \pm 0.02$   \\
ETASI 4  & 305887 & $2.26 \pm 0.01$  & $2.26 \pm 0.01$   & $2.68 \pm 0.07$  & $2.75 \pm 0.03$   \\
ETASI 5  & 730056 & $2.049 \pm 0.006$  & $2.153 \pm 0.006$  & $2.53 \pm 0.03$  & $2.71 \pm 0.03$  \\
ETASI 6  & 432091 & $2.131 \pm 0.008$  & $2.131 \pm 0.008$   & $2.57 \pm 0.06$  & $2.74 \pm 0.03$   \\
ETASI 7  & 840102 & $1.968 \pm 0.006$  & $2.043 \pm 0.006$  & $2.60 \pm 0.08$  & $2.73 \pm 0.02$   \\
ETASI 8  & 346611 & $1.964 \pm 0.006$  & $2.051 \pm 0.006$  & $2.55 \pm 0.05$  & $2.73 \pm 0.03$   \\
ETASI 9  & 335283 & $2.215 \pm 0.008$  & $2.215 \pm 0.008$   & $2.75 \pm 0.12$  & $2.75 \pm 0.03$   \\
ETASI 10 & 646730 & $2.049 \pm 0.006$  & $2.131 \pm 0.007$  & $2.52 \pm 0.04$  & $2.73 \pm 0.02$  \\
\hline
\end{tabular}
\caption{Comparison of different estimators for the $\beta$ value across five geographic regions and for ten different realizations of the ETASI catalog, with $\beta_{\text{true}} = 2.763$ implemented.}
\label{tab2}
\end{table}

\section{Discussion and Conclusions}

We have estimated the $b$-value in synthetic ETAS catalogs after the removal of small-magnitude earthquakes to account for STAI and SNDI. Our findings indicate that traditional methods, which rely on identifying a magnitude threshold $M_{th}$, above the completeness level, and fitting a GR law for magnitudes $m > M_{th}$, consistently underestimate the true $b$-value. In contrast, we show that the b-more positive estimator $\beta_{++}$ provides an accurate estimate of the true $b$-value. Applying this same analysis to instrumental catalogs, we observe that estimates from $\beta_{++}$ are consistently higher than those from traditional methods, suggesting that conventional $b$-value estimates may not fully capture the true $b$-value. A

A closer inspection of the results across regional catalogs (Figs. \ref{fig1} and \ref{fig3}) reveals a non-monotonic trend in $\beta_{++}$ as a function of $\delta M_{th}$ (or, equivalently, with decreasing $N_{th}$). This trend is clearly observable in Italy and especially in New Zealand, where $\beta_{++}$ decreases from a peak value of $\beta_{++} = 2.64 \pm 0.02$ at $\delta M_{th} = 0.5$ to $\beta_{++} = 2.55 \pm 0.03$ at $\delta M_{th} = 1.2$ for Italy, and from $\beta_{++} = 2.68 \pm 0.02$ at $\delta M_{th} = 0.3$ to $\beta_{++} = 2.53 \pm 0.03$ at $\delta M_{th} = 1.1$ for New Zealand. This decline is also noticeable, though to a lesser extent, in Japan and Southern California, while it is absent in Northern California. The initial increase in $\beta_{++}$ with $\delta M_{th}$ aligns with theoretical predictions (Sec.\ref{sec2.1}), and this trend is clearly reflected in synthetic ETASI catalogs (Fig. \ref{fig1}). However, the subsequent decrease in $\beta_{++}$ was not anticipated and may suggest a characteristic feature of real seismicity that the ETASI model does not fully capture. This behavior could indicate complex deviations from the GR law, potentially arising from correlations between earthquake magnitudes (\cite{LGdA07, LdAG08, NOS19, NOS22, LdAG24}). Clarifying whether the observed behavior of $\beta_{++}$ at higher values of $\delta M_{th}$ stems from statistical fluctuations or represents an intrinsic feature of seismicity remains a crucial challenge, with significant implications for improving seismic forecasting models.

\section*{Data Availability Statement}
Instrumental earthquake catalogs can be accessed online through the following links: \textbf{Japan}: \url{https://www.data.jma.go.jp/eqev/data/bulletin/index.html}, \textbf{Southern California}: \url{https://scedc.caltech.edu/data/alt-2011-dd-hauksson-yang-shearer.html}, \textbf{Northern California}: \url{https://ncedc.org/ncedc/catalog-search.html}, \textbf{Italy}: \url{https://horus.bo.ingv.it/}, and \textbf{New Zealand}: \url{https://quakesearch.geonet.org.nz/}. Additionally, tools for generating synthetic ETASI catalogs are available at \url{www.declustering.com}.

\section*{Acknowledgments}
 E.L. acknowledges support from the MIUR PRIN 2022 PNRR  P202247YKL, C.G. acknowledges support from the MIUR PRIN 2022 PNRR P20222B5P9, G.P. would like to acknowledge the Earth Observatory of Singapore (EOS), and the Singapore Ministry of Education Tier 3b project “Investigating Volcano and Earthquake Science and Technology (InVEST)”.

\bibliographystyle{apacite}    

\end{document}